\documentclass[aps,showpacs,11pt]{revtex4}
\usepackage{dcolumn}
\usepackage{graphicx}
\usepackage{amsmath}
\usepackage{amsfonts}
\usepackage{amssymb}
\usepackage{psfrag}
\usepackage{wrapfig}
\usepackage{subfigure}
\usepackage{makeidx}
\usepackage{bm}
\usepackage{epsf}
\usepackage{hyperref}
\usepackage{color}
\usepackage{slashed}

\begin{document}

\title{Holographic superconductor in hyperscaling violation geometry with Maxwell-dilaton coupling}

\author{Shao-Jun Zhang $^1$}
\email{sjzhang84@hotmail.com}
\author{Qiyuan Pan $^{1,2}$}
\email{panqiyuan@126.com}
\author{E. Abdalla $^1$}
\email{eabdalla@usp.br}
\affiliation{$^1$ Instituto de
F$\acute{i}$sica, Universidade de S$\tilde{a}$o Paulo, C.P. 66318,
05315-970, S$\tilde{a}$o Paulo, SP, Brazil\\
$^2$ Department of Physics, Key Laboratory of Low Dimensional Quantum Structures and Quantum Control of Ministry of Education, Hunan Normal University, Changsha, Hunan 410081, China}

\date{\today}

\begin{abstract}
We re-investigate the holographic superconductor in hyperscaling violation geometry by considering the coupling between the probed Maxwell field and the background dilaton. We find that the phenomenon of superconductivity still exists, but with properties affected by such a coupling. The critical temperature decreases as the hyperscaling violation exponent is increased. The influence of the dynamical exponent on the critical temperature becomes complicated which depends on the mass of the probed scalar field and the hyperscaling violation exponent. The results of the frequency gap show a large deviation from the expected universal relation.

\end{abstract}

\pacs{11.25.Tq, 04.70.Bw, 74.20.-z}

\keywords{Holographic superconductor; Hyperscaling violation; Maxwell-dilaton coupling; Gauge/gravity duality}

\maketitle

\section{Introduction}

The past decade has witnessed the great success of AdS/CFT duality~\cite{Maldacena:1997re,Gubser:1998bc,Witten:1998qj}, which relates a $(d+1)$-dimensional quantum field theory to a gravity theory living in $(d+2)$ dimensions. As a strong/weak duality, it has been successfully applied on various areas of modern physics to deal with problems where traditional methods confront great challenge or even break down. One such particular important application, the so-called AdS/CMT (condensed matter physics), is to gain a better understanding of numerous exotic but very important features of electronic materials, such as high temperature superconductors where BCS theory fails.  Within the standard framework of AdS/CFT, it was suggested that the normal/superconducting phase transition in boundary field theory corresponds to a second-order phase transition in the dual gravity side, where the bulk AdS black hole, under a charged scalar field perturbation, tends to develop a scalar hair~\cite{Gubser:2008px,Hartnoll:2008vx,Hartnoll:2008kx}. This so-called holographic superconductor model can indeed reproduce some properties of superconductor by doing calculations in the AdS black hole which is dual to relativistic CFT at finite temperature. Various holographic superconductor models have been developed, for reviews, see Refs.~\cite{Hartnoll:2009sz,Herzog:2009xv,Horowitz:2010gk,Cai:2015cya}.

Generally, systems near critical points exhibit a scaling symmetry and can be described by a CFT. However, many realistic condensed matter systems near critical points, although exhibiting a scaling symmetry, have different scalings in the spatial and time directions~\cite{Hartnoll:2009sz}. A prototype example of such critical points is a Lifshitz fixed point, where the degree of anisotropy between the spatial and time directions is parameterized by a dynamical exponent $z$. To describe such a Lifshitz fixed point in AdS/CFT, a non-relativistic version of this duality has been developed, where the bulk is a Lifshitz-like geometry rather than an asymptotically AdS geometry~\cite{Kachru:2008yh}. Many efforts have been devoted to construct holographic superconductors in these geometries~\cite{Brynjolfsson:2009ct,Sin:2009wi,Cai:2009hn,Momeni:2012tw,Schaposnik:2012cr,Bu:2012zzb,Abdalla:2011fd,
Abdalla:2013zra,Zhao:2013pva,Lu:2013tza,Tallarita:2014bga,Wu:2014dta,Lala:2014jca,
Lin:2014bya,Guo:2014wca,Jing:2014bza,Dector:2015cia}, and it is found that holographic superconductivity still exists. Many new features appear in these holographic superconductors. The effect of the dynamical exponent is explored~\cite{Lu:2013tza}, and it is found that, for fixed conformal dimension of the scalar operator, the critical temperature decreases as the dynamical exponent increases. The ratio of the gap frequency to the critical temperature is found to be a little larger than the one in the relativistic case~\cite{Cai:2009hn}.

Recently, the Lifshitz-like geometries are extended to even more sophisticated geometries, the so-called hyperscaling violating geometries~\cite{Charmousis:2010zz,Gouteraux:2011ce,Huijse:2011ef,
Dong:2012se,Cadoni:2012uf,Ammon:2012je,Alishahiha:2012cm,Alishahiha:2012qu,Salvio:2013jia,Ghodrati:2014spa}, which on top of anisotropic scaling, have also an overall hyperscaling factor under the scaling parameterized by the hyperscaling violation exponent $\theta$. That means the metric is no longer invariant under the scaling but covariant. The neutral hyperscaling violating geometries can be constructed starting from a Einstein-Maxwell-dilaton action~\cite{Huijse:2011ef,Dong:2012se,Alishahiha:2012cm}, where a $U(1)$ gauge field is coupled with a dilaton and responsible for generating the hyperscaling violation. Their charged cousins have also been constructed by adding into the action another $U(1)$ gauge field~\cite{Alishahiha:2012qu}, which is also required to couple with the dilaton to generate charged solutions. This kind of geometry has received considerable attention due to its potential applications in condensed matter physics~\cite{Fan:2013tpa,Fan:2013tga,Fan:2013zqa,Lucas:2014zea,Kuang:2014pna,Kuang:2014yya,Pan:2015a,Chen:2015azo}. An interesting feature of this kind of geometry appears when the hyperscaling violation exponent assumes the special value $\theta=d-1$, when the holographic entanglement entropy exhibits logarithmic violation of the usual area law~\cite{Huijse:2011ef}, indicating that this kind of geometry can provide a gravitational dual for the theory with a Fermi surface.

Holographic superconductors are also investigated in neutral black brane with hyperscaling violation to explore the effect of the hyperscaling violation exponent. In Refs.~\cite{Fan:2013tga,Pan:2015a}, it is found that the increase of the hyperscaling violation makes the condensation of the scalar operator harder for small $\theta$ but easier for large $\theta$. Moreover, the higher hyperscaling violation changes the usual relation in the gap frequency. In this holographic superconductor model, the probed Maxwell field is not coupled with the background dilaton. However, from the above statement, we know that this coupling is necessary to generate a charged hyperscaling violating black brane~\cite{Alishahiha:2012qu}. So it is natural to see if there is still superconductivity when such a coupling is taken into account. This is the main task of our present work. We will see that the superconductivity still exists but with features modified by such a coupling.

The work is organized as follows. In Sec.~II, we will briefly review the black brane background with hyperscaling violation. By applying both numerical and Sturm-Liouville (S-L) analytical methods, we will explore the effect of the hyperscaling violation exponent on the normal/superconducting phase transition. In Sec.~III, we discuss the conductivity and the influence of the hyperscaling violation. The last section is devoted to a summary and to conclusions.

\section{Black hole solution with hyperscaling violation}

The black brane solution with hyperscaling violation can be obtained from the following Einstein-Maxwell-dilaton action~\cite{Dong:2012se,Alishahiha:2012qu}
\begin{eqnarray}
S = - \frac{1}{16\pi G} \int d^{d+2}x \sqrt{-g} \left[R-\frac{1}{2} (\partial \tilde{\phi})^2 + V(\tilde{\phi}) -\frac{1}{4} e^{\lambda_1 \tilde{\phi}} {\cal F}_{\mu\nu} {\cal F}^{\mu\nu}\right],\label{action}
\end{eqnarray}
where the scalar potential takes the form
\begin{eqnarray}
V=V_0 e^{\gamma \tilde{\phi}},
\end{eqnarray}
with $V_0$ a constant. The gauge field ${\cal F}_{\mu\nu} \equiv \partial_\mu {\cal A}_\nu - \partial_\nu {\cal A}_\mu$ is responsible for generating hyperscaling violation. By solving the equations of motion derived from the above action, we have the following black brane solution with hyperscaling violation,
\begin{eqnarray}
ds^2 &=& r^{-2\theta/d} \left[-r^{2z} f(r) dt^2 + \frac{dr^2}{f(r)} + r^2 d\vec{x}^2\right],\nonumber\\
f(r) &=& 1-\left(\frac{r_H}{r}\right)^{d+z-\theta},\nonumber\\
{\cal A} &=& -\slashed{\mu} r_H^{d+z-\theta} \left[1-\left(\frac{r}{r_H}\right)^{d+z-\theta}\right] dt,\nonumber\\
e^{\tilde{\phi}} &=& r^{\beta},
\end{eqnarray}
where $\theta$ is the hyperscaling violation exponent~\cite{Fisher:1986zz}, $z$ is the dynamical exponent and $r_H$ is the event horizon. The other parameters are given by the following relations
\begin{eqnarray}
\lambda_1 &=& - \frac{2(d-\theta+\theta/d)}{\sqrt{2(d-\theta)(z-1-\theta/d)}},\nonumber\\
\gamma &=& \frac{2\theta/d}{\sqrt{2(d-\theta)(z-1-\theta/d)}},\nonumber\\
V_0 &=& (d+z-\theta-1)(d+z-\theta),\nonumber\\
\slashed{\mu} &=& \sqrt{\frac{2(z-1)}{d+z-\theta}},\nonumber\\
\beta &=& \sqrt{2(d-\theta)(z-1-\theta/d)}.
\end{eqnarray}
This geometry has the following scaling behavior,
\begin{eqnarray}
t \rightarrow \alpha^z t,\quad r \rightarrow \alpha^{-1} r,\quad x_i \rightarrow \alpha x_i,\quad ds\rightarrow \alpha^{\theta/d} ds,
\end{eqnarray}
with a real positive number $\alpha$. It is interesting to note that the line element is not invariant under the scaling but covariant with the hyperscaling violation exponent $\theta$.

For convenience, in the following discussion, we introduce a dimensionless radial coordinate $u \equiv r_H/r$ and rewrite the metric and the dilaton as
\begin{eqnarray}\label{geometry}
ds^2 &=& \left(\frac{u}{r_H}\right)^{-2(1-\theta/d)} \left[-\left(\frac{u}{r_H}\right)^{-2(z-1)} f(u) dt^2 +\frac{du^2}{r_H^2 f(u)} +d \vec{x}^2\right],\nonumber\\
e^{\tilde{\phi}} &=& \left(\frac{u}{r_H}\right)^{-\beta}.
\end{eqnarray}
with $f(u)=1-u^\Delta$ and $\Delta \equiv {d+z-\theta}$. With this coordinate, $u=1$ corresponds to the horizon and $u=0$ to the boundary at infinity. The Hawking temperature of the black hole is
\begin{eqnarray}
T=\frac{\Delta r_H^z}{4\pi},
\end{eqnarray}
which is interpreted as the temperature of the dual field theory. The hyperscaling violation exponent $\theta$ is constrained as~\cite{Dong:2012se}
\begin{eqnarray}
d>\theta \geq 0,\quad z \geq 1+\frac{\theta}{d}.
\end{eqnarray}
We will fix $d=2$ in the following sections, this means that the dual field theory is $(2+1)$-dimensional.

\section{Holographic superconductor with probed Maxwell field coupled with dilaton}

To study the holographic superconductor with hyperscaling violation in the probe limit, we consider the Maxwell-scalar action
\begin{eqnarray}
S = \int d^{d+2}x \sqrt{-g} \left(-\frac{1}{4} e^{\lambda_2 \tilde{\phi}} F_{\mu\nu} F^{\mu\nu} - |\nabla \psi - i e A \psi|^2 -m^2 |\psi|^2\right).
\end{eqnarray}
In contrast to refs.~\cite{Fan:2013tga,Pan:2015a}, here we consider the probed Maxwell field $F_{\mu\nu} \equiv \partial_\mu A_\nu -\partial_\nu A_\mu$ to be coupled with the background dilaton. The coupling constant is $\lambda_2 = \sqrt{\frac{2(z-1-\theta/d)}{d-\theta}}$. The motivation of considering such a coupling comes from the charged black brane solution with hyperscaling violation~\cite{Alishahiha:2012qu}, where such a coupling is required to generate an analytical solution. The existence of this coupling has a large influence on the condensation, as we will see in the following. It should be noted that the gauge field in Eq.~(\ref{action}) is simply used to generate the hyperscaling violation and remains unbroken throughout this work. $e$ is the charge the scalar field carries and we set it to unity without loss of generality.

The equations of motion are
\begin{eqnarray}
&&\nabla_\mu \left(e^{\lambda_2 \tilde{\phi}} F^{\mu\nu}\right) = i e (\psi^\ast \nabla^\nu \psi-\psi \nabla^\nu \psi^\ast) + 2 e^2 A^\nu |\psi|^2,\nonumber\\
&&\nabla_\mu \nabla^\mu \psi-2 i e A^\mu \partial_\mu \psi -(e^2 A_\mu A^\mu+m^2) \psi=0.
\end{eqnarray}
Taking the usual ansatz $\psi = \psi(u)$ and $A=\phi(u) dt$, the equations of motion can be rewritten as
\begin{eqnarray}
\psi'' + \left(\frac{f'}{f}-\frac{\Delta-1}{u}\right)\psi' + \left[\frac{e^2 r_H^{-2z} u^{2(z-1)} \phi^2}{f^2}- \frac{m^2 r_H^{-2\theta/d}}{u^{2(1-\theta/d)} f} \right]\psi=0,\label{psiEq}\\
\phi'' + \frac{3-\Delta}{u}\phi'- \frac{2 e^2 r_H^{2(1-z)} \psi^2}{u^{4-2z} f}\phi=0.\label{phiEQ}
\end{eqnarray}
Regularity at the horizon ($u=1$) demands the boundary conditions
\begin{eqnarray}
\psi(1) = - \frac{\Delta}{m^2 r_H^{-2\theta/d}} \psi'(1),\qquad \phi(1)=0. \label{BnyHorizon}
\end{eqnarray}
At the infinite boundary ($u\rightarrow 0$), the solutions have the asymptotic behavior
\begin{eqnarray}
\psi = \Bigg\{\begin{array}{ll}
\psi_-r_H^{-\Delta_-} u^{\Delta_-} + \psi_+ r_H^{-\Delta_+} u^{\Delta_+},\quad & \mathrm{with}\ \Delta_\pm =\frac{\Delta}{2} \pm \sqrt{\left(\frac{\Delta}{2}\right)^2 + m^2 r_H^{-2\theta/d}} \ \mathrm{for} \ \theta=0,\\
\psi_-+ \psi_+ r_H^{-\Delta_+} u^{\Delta_+},\quad & \mathrm{with} \ \Delta_+ =\Delta \ \mathrm{for} \ \theta\neq 0,
\end{array}\label{psiBny}
\end{eqnarray}
\begin{eqnarray}
\phi = \mu + \rho r_H^{2-\Delta} u^{\Delta-2}.\label{phiBny}
\end{eqnarray}
According to the recipe of the gauge/gravity duality, $\Delta_+$ is the conformal dimension of the scalar operator $\langle O \rangle =\psi_+$ which is dual to the bulk scalar field. $\mu$ and $\rho$ are interpreted as the chemical potential and charge density in the dual field theory respectively. We will focus on the condensation of the operator $\langle O \rangle$ and set the source $\psi_-=0$, as done in Refs.~\cite{Fan:2013tga,Pan:2015a}.

\subsection{Numerical investigation}

In this section, we will solve Eqs.~(\ref{psiEq}) and (\ref{phiEQ}) numerically with shooting method. Before presenting our results, it is useful to note that the equations of motion have the scaling symmetry
\begin{eqnarray}\label{scaling}
&&\psi \rightarrow \alpha^{1-z} \psi,\quad \phi \rightarrow \alpha^{-z} \phi,\nonumber\\
&&\psi_+ \rightarrow \alpha^{1-\Delta_+ -z} \psi_+,\quad \mu \rightarrow \alpha^{-z} \mu.
\end{eqnarray}

Choosing various dynamical exponent $z$ and hyperscaling violation exponent $\theta$, we solve Eqs.~(\ref{psiEq}) and (\ref{phiEQ}) numerically with the boundary conditions Eqs.~(\ref{psiBny}) and (\ref{phiBny}), from which we can read off the value of $\psi_+$ which gives the condensation $\langle O \rangle$. In Fig.~1, we show the results for two different dynamical exponents $z=1.5$ (left) and $z=2$ (right), respectively. From the figure, we can see the condensation, which is continuous near the critical temperature and dropping to zero at the critical point. By fitting the curves near the critical point, we have the relation
\begin{eqnarray}
\langle O \rangle \sim \left(1-T/T_c\right)^{1/2},\label{CondenScaling}
\end{eqnarray}
which is typical of second order phase transition with the mean-field exponent $1/2$. We will confirm this result by using analytical method later. Our results show that there should still exist superconductivity when we consider the coupling between the probed Maxwell field and the background dilaton.

\begin{figure}[!htbp]
\centering
\subfigure[~~$z=1.5$]{\includegraphics[width=0.48\textwidth]{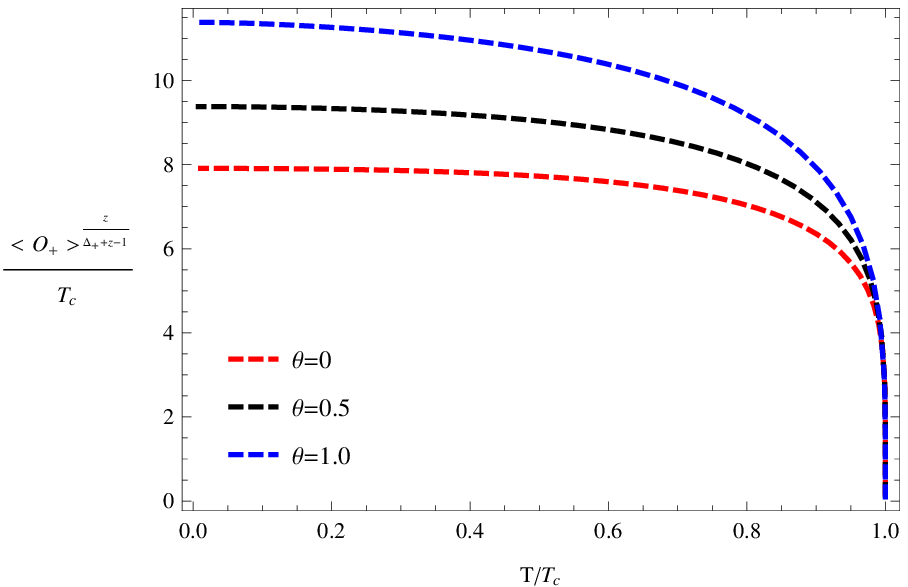}}\quad
\subfigure[~~$z=2$]{\includegraphics[width=0.48\textwidth]{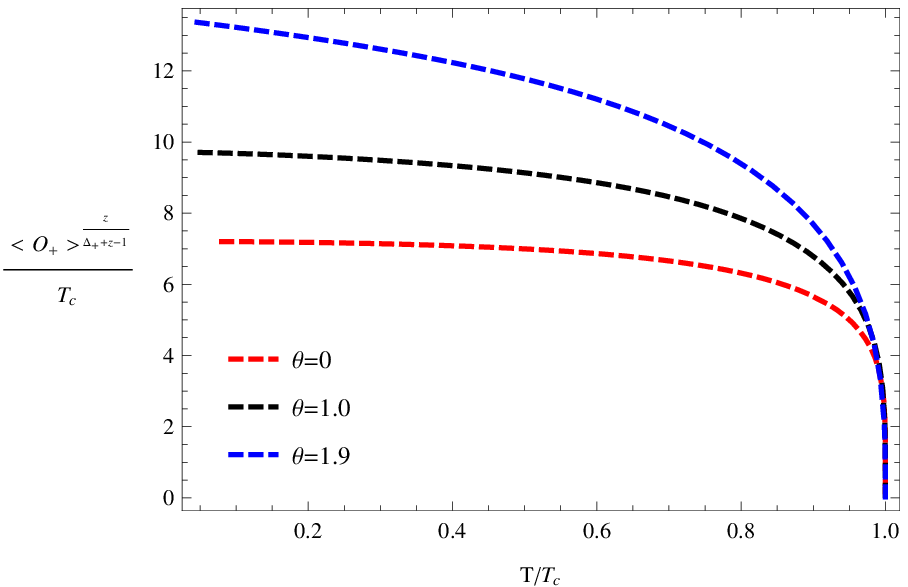}}\quad
\caption{(Color online) The condensate of the scalar operator $\langle O \rangle$ as a function of temperature for two different dynamical exponents $z=1.5$ (left) and $z=2$ (right). The mass of the scalar field is fixed to be $m^2 r_H^{-2\theta/d}=0$. In each panel, the three lines correspond to different values of the hyperscaling violation exponent.}
\end{figure}

To see the influence of the dynamical exponent $z$ and the hyperscaling violation exponent $\theta$ on the condensation, we calculate the critical temperature for various values of $z$ and $\theta$. In Fig.~2, we show some of the results. We can see that, for fixed $z$ and the mass of the scalar field,  when we increase the hyperscaling violation exponent $\theta$, the critical temperature decreases. This result is completely different from the case without the coupling between the Maxwell field and the backgound dilaton, where, as $\theta$ increases, the critical temperature first decreases and then increases~\cite{Pan:2015a}.

\begin{figure}[!htbp]
\centering
\subfigure[~~$m^2 r_H^{-2\theta/d}=0$]{\includegraphics[width=0.32\textwidth]{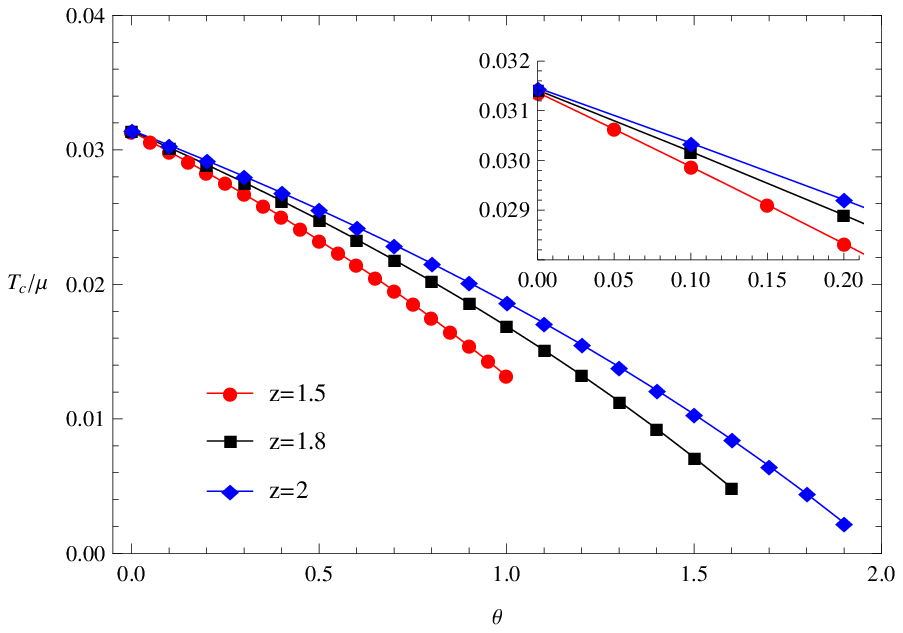}}
\subfigure[~~$m^2 r_H^{-2\theta/d}=-1$]{\includegraphics[width=0.32\textwidth]{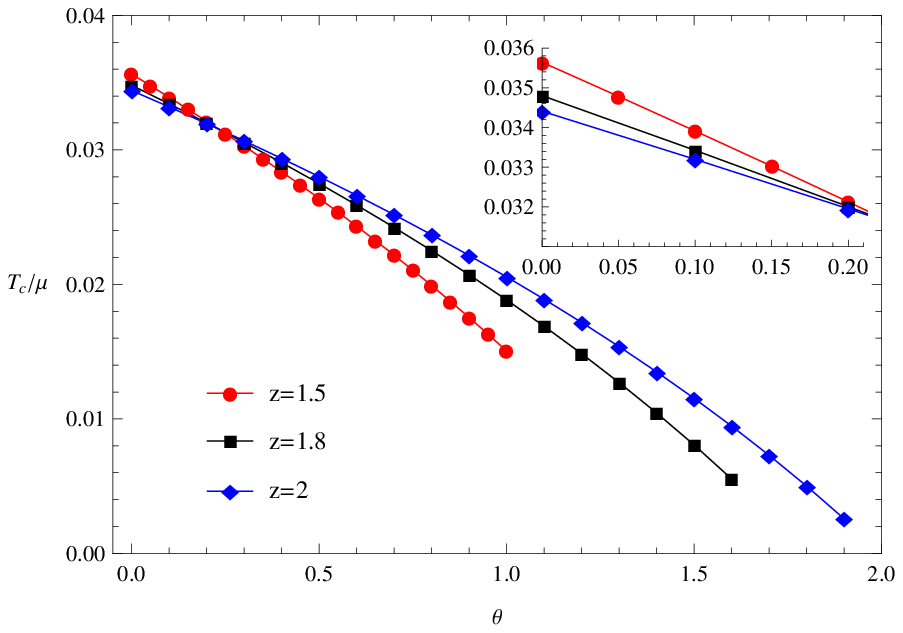}}
\subfigure[~~$m^2 r_H^{-2\theta/d}=-3$]{\includegraphics[width=0.32\textwidth]{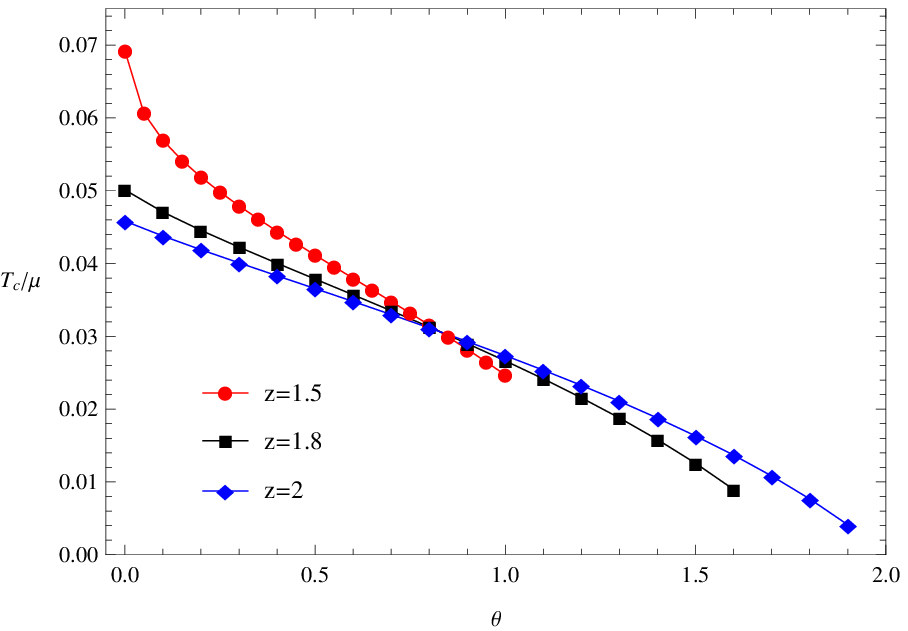}}
\caption{(Color online) The critical temperature as a function of the hyperscaling violation $\theta$ for three different masses of the scalar field. In each panel, the three different lines correspond to different dynamical exponent $z$, i.e., $z=1.5$ (red), $z=1.8$ (black), and $z=2$ (blue).}
\end{figure}

We can also see that, for fixed $\theta$ and the mass of the scalar field, the critical temperature is not always a monotonic function of the dynamical exponent $z$, the situation depends both on the mass of the scalar field and the hypercaling violation exponent $\theta$. From the left panel where $m^2 r_H^{-2\theta/d}=0$, we can see that the critical temperature increases as $z$ increases, even for $\theta=0$ ($T_c /\mu= 0.031368, 0.031415, 0.031459$ for $z=1.5, 1.8, 2.0$, respectively). However, when we decrease the mass of the scalar field, the critical temperature is no longer a monotonic function of $z$. From the last two panels where $m^2 r_H^{-2\theta/d}=-1$ and $-3$ respectively, we can see that the three lines with different $z$ intersect at a point $\theta=\theta_c$. When $0 \leq \theta<\theta_c$, the critical temperature decreases as $z$ increases, while for $\theta>\theta_c$ we have opposite behavior.
It should be noted that the value of $\theta_c$ depends on the mass of the scalar field.

\subsection{Analytical investigation}

In this section, we use the S-L analytical method~\cite{Siopsis:2010uq} to analyze the critical temperature and critical phenomena, and then to confirm our numerical results derived above.

At the critical temperature $T=T_c$, there is no condensation $\psi=0$, thus the equation of motion of $\phi$ ~(\ref{phiEQ}) reduces to
\begin{eqnarray}
\phi''+\frac{3-\Delta}{u} \phi'=0.
\end{eqnarray}
With the boundary condition~(\ref{BnyHorizon}), we obtain the solution
\begin{eqnarray}
\phi = \lambda  r_{Hc}^z (1-u^{\Delta-2}),\qquad \lambda \equiv \mu r_{Hc}^{-z},
\end{eqnarray}
where $r_{Hc}$ denotes the horizon radius at the critical point.

With the boundary condition~(\ref{psiBny}) we assume $\psi$ takes the form
\begin{eqnarray}
\psi (u) \sim \langle O \rangle r_H^{-\Delta_+} u^{\Delta_+} F(u),
\end{eqnarray}
where the trial function $F(u)$ satisfies the boundary conditions $F(0)=1$ and $F'(0)=0$. Then the equation of motion for $\psi$, Eq.~(\ref{psiEq}), can be written in the form
\begin{eqnarray}
(Q F')' + Q(U+\lambda^2 V)F=0,\label{SLeq}
\end{eqnarray}
with
\begin{eqnarray}
Q=u^{2\Delta_+ - \Delta +1}f,\quad U=\frac{\Delta_+ f'}{u f}+\frac{\Delta_+ (\Delta_+ - \Delta)}{u^2} -\frac{m^2 r_H^{-2\theta/d}}{u^{2(1-\theta/d)} f},\quad V=\frac{e^2 u^{2(z-1)} (1-u^{\Delta-2})^2}{f^2}.
\end{eqnarray}
This takes the standard form of the Sturm-Liouville equation~\cite{Gelfand:1963} and the minimal eigenvalue of $\lambda$ can be estimated by minimizing the expression
\begin{eqnarray}
\lambda^2 = \frac{\int_0^1 Q(F'^2-U F^2) du}{\int_0^1 Q V F^2 du}.
\end{eqnarray}
After getting $\lambda_{min}$ we can calculate the critical temperature as
\begin{eqnarray}
T_c = \frac{\Delta}{4\pi \lambda_{min}} \mu.
\end{eqnarray}

The trial function $F(u)$ can be chosen as a polynomial function satisfying the boundary conditions, for example we can choose
\begin{eqnarray}
F(u) = 1-a u^2,
\end{eqnarray}
with a constant $a$ to be determined by minimizing the expression. Also, we can choose $F(u)$ to contain higher order terms as done in Ref.~\cite{Pan:2015a}. But in our case, second order is enough to re-produce the numerical results.

\begin{figure}[!htbp]
\centering
\subfigure[~~$m^2 r_H^{-2\theta/d}=0$]{\includegraphics[width=0.45\textwidth]{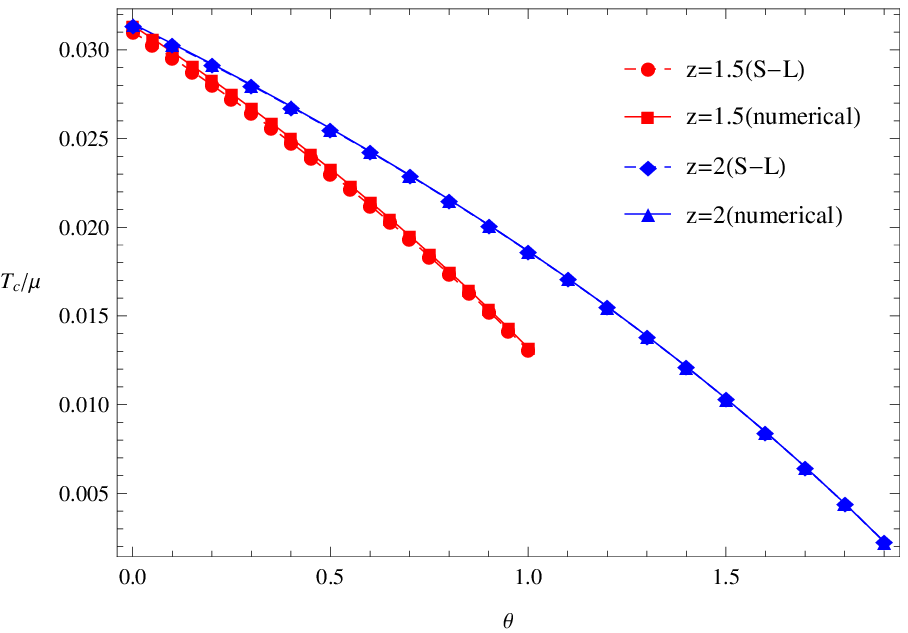}}\quad
\subfigure[~~$m^2 r_H^{-2\theta/d}=-1$]{\includegraphics[width=0.45\textwidth]{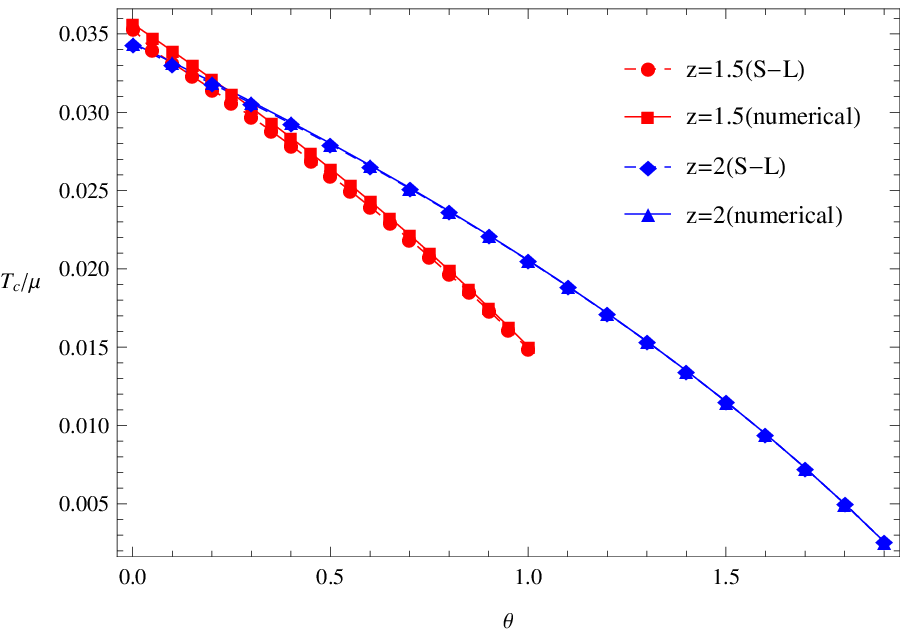}}
\caption{(Color online) The critical temperature as a function of the hyperscaling violation $\theta$ derived from S-L analytical method (dashed lines) and numerical method (solid lines). The mass of the scalar field is fixed to be $m^2 r_H^{-2\theta/d}=0$ (left) and $m^2 r_H^{-2\theta/d}=-1$ (right), respectively. In each panel, the dynamical exponent takes two values, i.e., $z=1.5$ (red) and $z=2$ (blue).}
\end{figure}

In Fig.~3, we show the relation between the critical temperature and the hyperscaling violation exponent $\theta$ derived from the S-L method. We also show the numerical results here derived above to make a comparison. From the figure, we can see that, for the cases we considered, the results derived from the S-L method match very well with the numerical results.

Next let us analyze the critical phenomena near the critical temperature. As the condensation $\langle O \rangle$ is very small at this moment, we can expand $\phi(u)$ in $\langle O \rangle$ as
\begin{eqnarray}
\phi(u) = \lambda r_H^z (1-u^{\Delta -2}) + r_H^z {\cal A} \chi(u) + \cdots,
\end{eqnarray}
where we have introduced ${\cal A} \equiv r_H^{-2(\Delta_+ + z -1)} \langle O \rangle^2$ and $\chi(u)$ satisfies the boundary conditions $\chi(1)=0$ and $\chi'(1)=0$~\cite{Siopsis:2010uq}. Then the $\phi$-equation~(\ref{phiEQ}) becomes
\begin{eqnarray}
(u^{3-\Delta} \chi')' = \frac{2 \lambda e^2 u^{2\Delta_+ - \Delta +2 z-1} (1-u^{\Delta-2}) F^2}{f}.
\end{eqnarray}
Doing one step of integration on both sides, we get
\begin{eqnarray}
(u^{3-\Delta} \chi')|_{u\rightarrow 0} = \lambda {\cal C},\qquad {\cal C} \equiv - \int_0^1 \frac{2 \lambda e^2 u^{2\Delta_+ - \Delta +2 z-1} (1-u^{\Delta-2}) F^2}{f} du.
\end{eqnarray}
From the above expression, we know that $\chi'(u) \sim u^{\Delta-3}$ as $u\rightarrow 0$.

On the other hand, near the critical temperature $\phi(u) \approx \mu (1-u^{\Delta-2})$. Thus at the infinite boundary $u\rightarrow 0$, we have
\begin{eqnarray}
\mu (1-u^{\Delta-2}) = \lambda r_H^z (1-u^{\Delta-2}) + r_H^z {\cal A} \left[\chi(0)+\chi'(0) u+ \cdots\right].
\end{eqnarray}
By equating the $u^{\Delta-2}$ term of both sides, we get
\begin{eqnarray}
\mu = \lambda r_H^z + r_H^z {\cal A} (u^{3-\Delta} \chi')|_{u\rightarrow 0},
\end{eqnarray}
leading to
\begin{eqnarray}
\langle O \rangle = \frac{1}{\sqrt{{\cal C}}} \left(\frac{4\pi T_c}{\Delta}\right)^{\frac{\Delta_+ +z -1}{z}} \left(1-\frac{T}{T_c}\right)^{1/2}.
\end{eqnarray}
The critical exponent assumes the mean-field value $\frac{1}{2}$. This confirms our previous result~(\ref{CondenScaling}).

\section{Conductivity}

To compute the conductivity, we consider a perturbed Maxwell field in $x$-direction taking the form as $\delta A_x = A_x(u) e^{-i \omega t} dx$. The equation of motion for $A_x(u)$ is
\begin{eqnarray}
A_x'' + \left(\frac{f'}{f}-\frac{d+3 z -\theta-5}{u}\right) A_x' + \left[\frac{r_H^{-2 z} \omega^2 u^{2(z-1)}}{f^2}-\frac{2 e^2 r_H^{2(1-z)} \psi^2}{u^{4-2 z} f}\right] A_x=0.
\end{eqnarray}
Near the horizon $u \rightarrow 1$, we take the ingoing boundary condition
\begin{eqnarray}
A_x (u) \sim (1-u)^{-\frac{i \omega}{4\pi T}},
\end{eqnarray}
and at infinite boundary $(u\rightarrow 0)$ the behavior is
\begin{eqnarray}
A_x = \Bigg\{
\begin{array}{ll}
A_x^{(0)}+A_x^{(1)} r_H^{-\Delta_x} u^{\Delta_x},\quad & \mathrm{with}\ \Delta_x=d+3 z -\theta-4 \ \mathrm{for}\ z-\theta \neq 2,\\
A_x^{(0)}-A_x^{(0)} \frac{\omega^2 r_H^{-2z} u^{2z} \ln u}{2 z} +A_x^{(1)} r_H^{-2z} u^{2z},\quad & \mathrm{for}\ z-\theta = 2\  (d=2).
\end{array}
\end{eqnarray}
It should be noted that a logarithmic term will appear in the asymptotic form of $A_x$ when $z-\theta=2$ for $d=2$. Then the conductivity of the dual superconductor, after suitable renormalization, can be calculated as~\cite{Hartnoll:2008vx,Hartnoll:2008kx,Kuang:2015mlf}
\begin{eqnarray}
\sigma = \Bigg\{
\begin{array}{ll}
- \frac{ i \Delta_x A^{(1)}}{\omega A^{(0)}}, & \mathrm{for}\ z-\theta \neq 2,\\
-\frac{2 i z A^{(1)}}{\omega A^{(0)}} + \frac{i \omega}{2 z}, & \mathrm{for}\ z-\theta = 2,
\end{array}
\end{eqnarray}
which transforms as $\sigma \rightarrow \alpha^{z-\Delta_x} \sigma$ under scaling.

\begin{figure}[!htbp]
\centering
\includegraphics[width=0.32\textwidth]{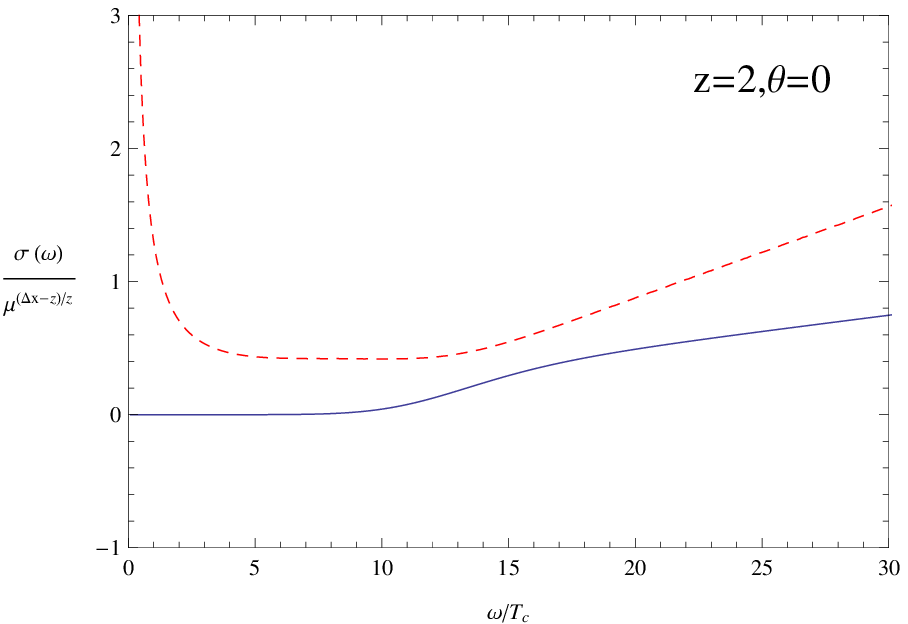}
\includegraphics[width=0.32\textwidth]{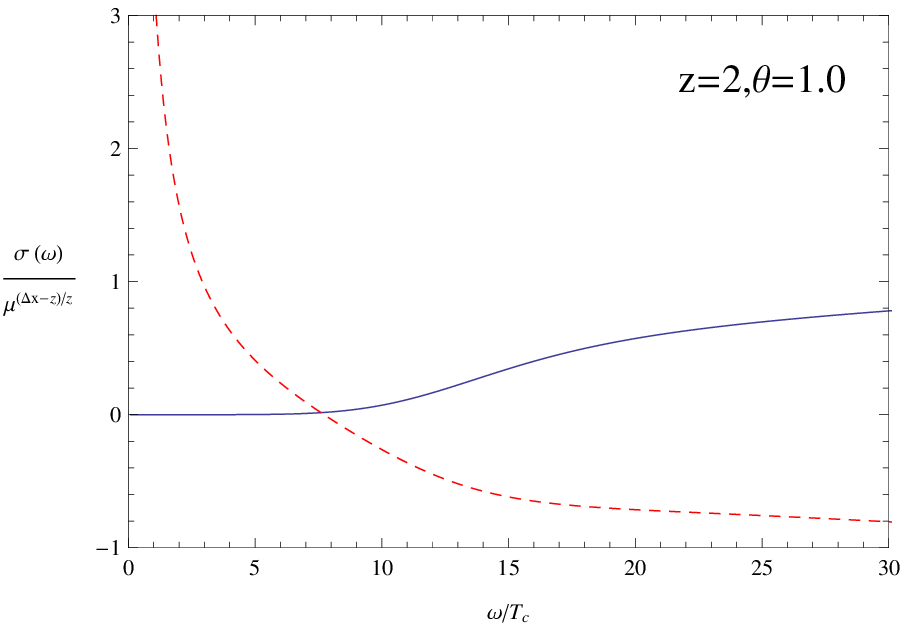}
\includegraphics[width=0.32\textwidth]{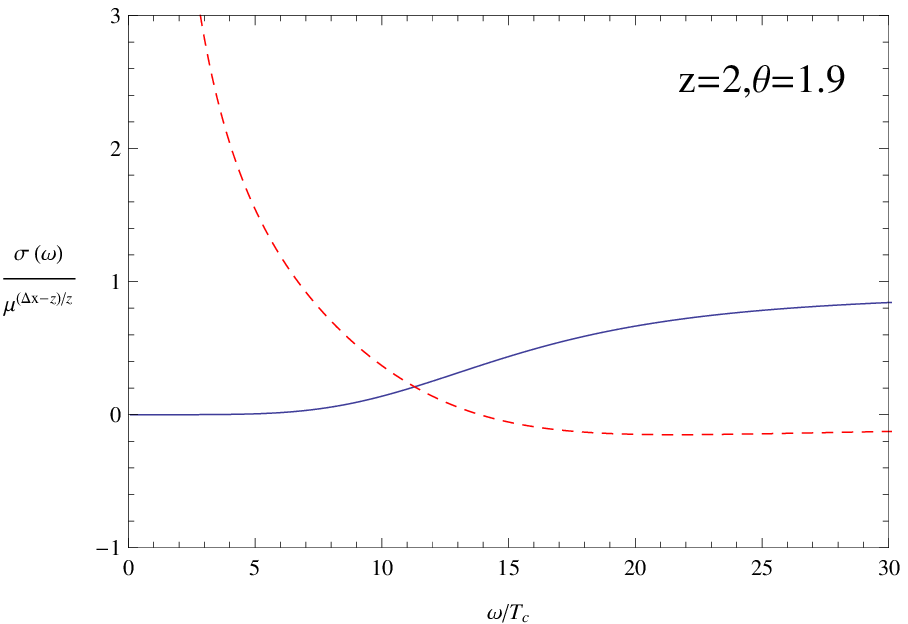}
\includegraphics[width=0.32\textwidth]{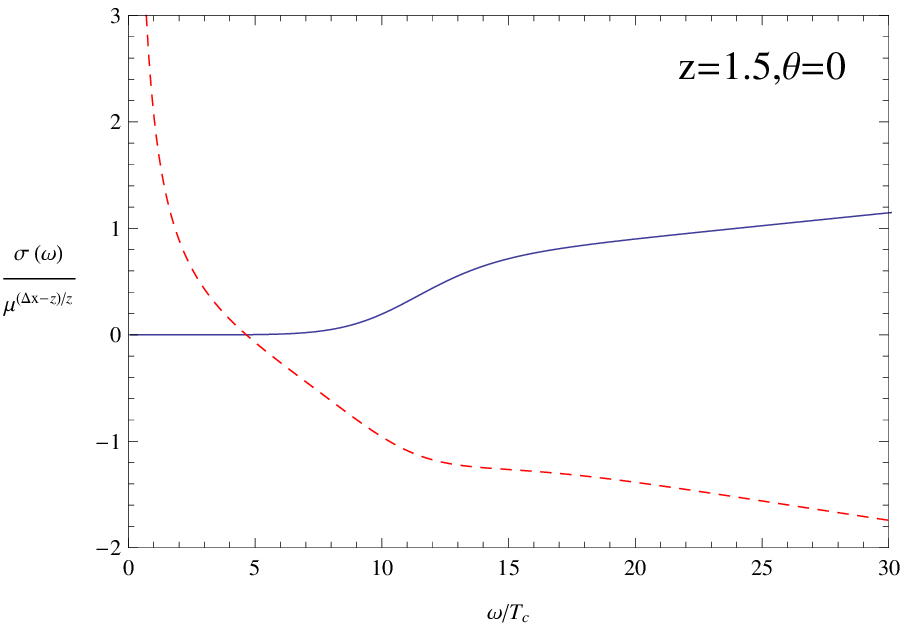}
\includegraphics[width=0.32\textwidth]{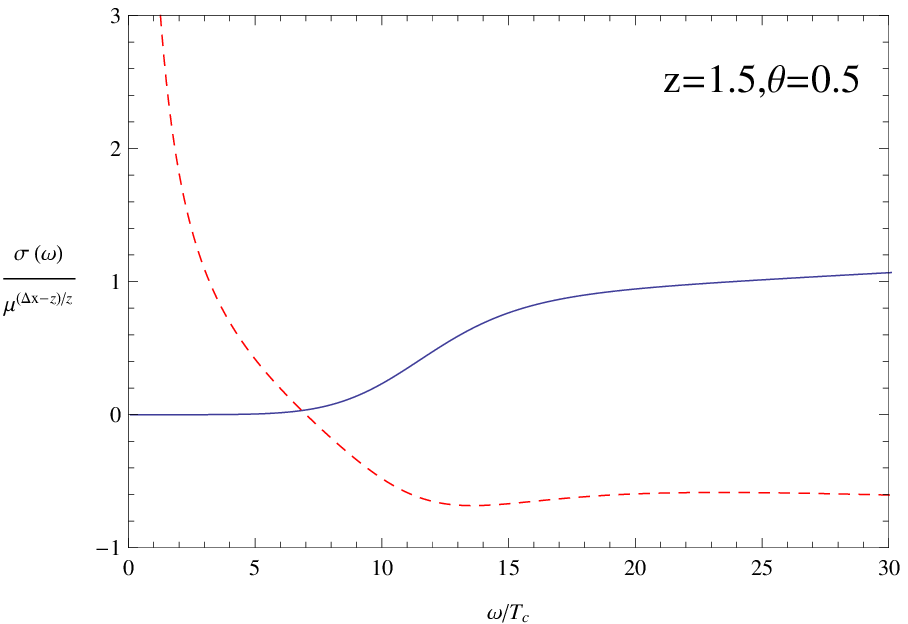}
\includegraphics[width=0.32\textwidth]{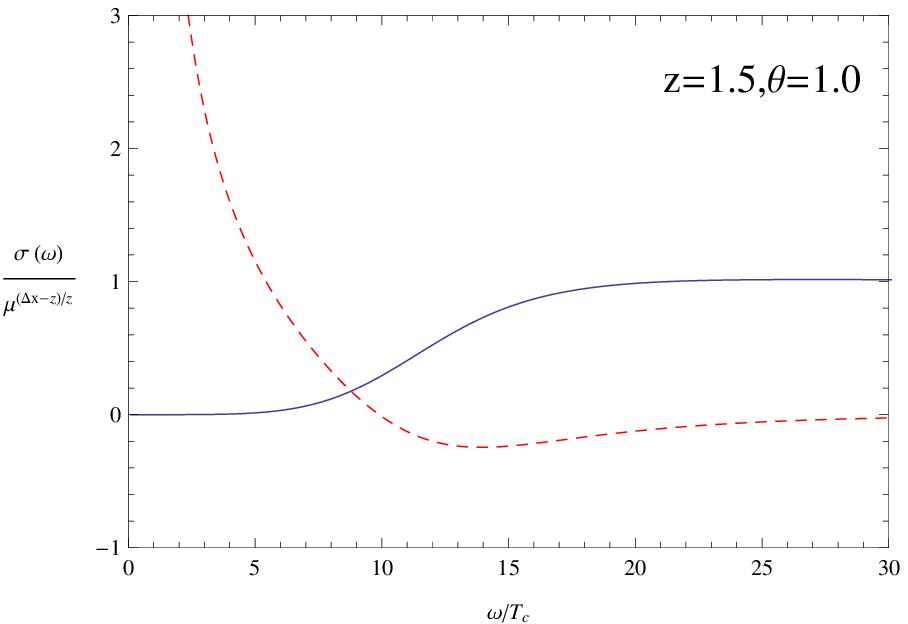}
\caption{(Color online) Conductivity of the holographic superconductor with hyperscaling violation at $T/T_c \approx 0.15$. The mass is fixed to be $m^2 r_H^{-2\theta/d}=0$. In each panel, the solid (blue) line and dashed (red) line represent the real and imaginary part of the conductivity, respectively.}
\end{figure}

In Fig.~4, we plot the frequency dependent conductivity of the holographic superconductor at temperature $T/T_c \approx 0.15$. From the figure, we can see that the there is a pole in the imaginary part of the conductivity, which indicates superconductivity. Compared to the results in the case without the Maxwell-dilaton coupling~\cite{Pan:2015a}, the curves show some different behaviors, especially for small $\theta$. From the panels with $\theta=0$, we can see that the imaginary part of the conductivity does not tend to be a constant as the frequency increases but diverges linearly. Similar phenomenon has been observed in Ref.~\cite{Lu:2013tza} for holographic superconductor in Lifshitz spacetime. The physical meaning of this phenomenon is not clear at present and needs further explanations.

From the figure, as in Refs.~\cite{Fan:2013tga,Pan:2015a}, we also find a gap in the superconductivity which is parameterized by the gap frequency $\omega_g$. Defining $\omega_g$ by the minimum of $|\sigma|$~\cite{Horowitz:2008bn}, we observe that for fixed $z$ the frequency gap $\omega_g$ increases as we increase $\theta$. This behavior is the same as in the case without the Maxwell-dilaton coupling~\cite{Pan:2015a}. By comparing results for different $z$ and fixed $\theta$, we can see that the gap $\omega_g$ decreases as we increase $z$. Moreover, compared to Ref.~\cite{Pan:2015a}, when taking into account the Maxwell-dilaton coupling, the frequency gap shows an even larger deviation from the relation $\omega_g/T_c \approx 8$~\cite{Horowitz:2008bn}, especially when $\theta$ is very small or very large.

\section{Summary and conclusions}

In this work, we discussed the holographic superconductor in a hyperscaling violating black brane. Different from previous work~\cite{Fan:2013tga,Pan:2015a}, the probed Maxwell field we considered is coupled with the background dilaton. The motivation for considering such a coupling comes from charged black brane solution with hyperscaling violation~\cite{Alishahiha:2012qu}, where such a coupling is necessary to generate an analytical solution. We find that when such a coupling is involved, features of the superconductor are changed. As we increase the hyperscaling violation exponent $\theta$, the critical temperature decreases, which means larger hyperscaling violation makes condensation of the dual scalar operator harder. This result is completely different from the one in case without such coupling, where the increase of the hyperscaling violation exponent makes the condensation harder for small $\theta$ but easier for large $\theta$. The influence of the dynamical exponent $z$ on the condensation is more complicated, depending on the mass of the scalar field $m$ and the hyperscaling violation exponent $\theta$. Our results show that when $m^2 r_H^{-2\theta/d}=0$, the critical temperature increases as $z$ increases, even for $\theta=0$. However, when we decrease the mass of the scalar field, the relation between $T_c$ and $z$ is no longer so simple. There exists a point $\theta=\theta_c$ where $T_c$ for different $z$ coincides. For $0 \leq \theta<\theta_c$, $T_c$ decreases as $z$ increases; however, for $\theta>\theta_c$, we have the opposite behavior. The critical behavior is also discussed and it is interesting to see that the critical exponent of the condensation still takes a mean-field value $1/2$. We confirm our numerical results by applying the analytical S-L method.

This coupling also affects the conductivity. When $\theta$ is very small, our results show that the imaginary part of the conductivity does not tend to be a constant but diverges linearly. The physical meaning of this exotic phenomenon is not clear at present and needs further explanations. There still exists a gap frequency but the expected universal relation $\omega_g/T_c \approx 8$ is changed, especially for very small or very large hyperscaling violation exponent.

The results coming out from the two holographic superconductor models, with and without the Maxwell-dilaton coupling, are so different, that they may be used to distinguish the two models. Here we only discuss s-wave holographic superconductor. It is interesting to see if such differences caused by this coupling still exist in p-wave and d-wave cases, and in insulator/superconductor phase transition. We leave it for further investigations.

\section*{Acknowledgement}

We would like to thank Xiao-Mei Kuang for helpful discussions. This work has been supported by CNPq No.~150035/2015-2 and FAPESP No.~2013/26173-9; the National Natural Science Foundation of China under grant No.~11275066; and the Open Project Program of State Key Laboratory of Theoretical Physics, Institute of Theoretical Physics, Chinese Academy of Sciences, China (No.~Y5KF161CJ1).

\end{document}